\documentclass[amsmath, amssymb, preprintnumbers, showpacs, showkeys, aps,prl,superscriptaddress,twocolumn]{revtex4-1}
\usepackage{graphicx}
\usepackage{calc} 
\usepackage{braket}
\usepackage{ulem}   
\usepackage{slashed}
\usepackage{wasysym}
\usepackage{dsfont}
\usepackage{amsthm,amsmath,amsfonts,amssymb,verbatim,color}
\usepackage{graphicx}
\usepackage{subfigure}
\usepackage{bm}
\usepackage{epsfig,slashed}
\usepackage[T1]{fontenc}
\usepackage[colorlinks=true,citecolor=blue,linkcolor=blue,urlcolor=blue]{hyperref}
\renewcommand{\b}[1]{{\mathbf{#1}}}

\begin{document}
\preprint{}

\title{Fermionic Symmetry-Protected Topological Phase in a Two-dimensional Hubbard Model}

%
%

\author{Cheng-Chien Chen}
\affiliation{Advanced Photon Source, Argonne National Laboratory, Argonne, Illinois 60439, USA}
\affiliation{Department of Physics, University of Alabama at Birmingham, Birmingham, Alabama 35294, USA}
\author{Lukas Muechler}
\affiliation{Department of Chemistry, Princeton University, Princeton, New Jersey 08544, USA}
\author{Roberto Car}
\affiliation{Department of Chemistry, Princeton University, Princeton, New Jersey 08544, USA}
\author{Titus Neupert}
\affiliation{Princeton Center for Theoretical Science, Princeton University, Princeton, New Jersey 08544, USA}
\author{Joseph Maciejko}
\affiliation{Department of Physics, University of Alberta, Edmonton, Alberta T6G 2E1, Canada}
\affiliation{Theoretical Physics Institute, University of Alberta, Edmonton, Alberta T6G 2E1, Canada}
\affiliation{Canadian Institute for Advanced Research, Toronto, Ontario M5G 1Z8, Canada}

\date\today

\begin{abstract}
We study the two-dimensional (2D) Hubbard model using exact diagonalization for spin-1/2 fermions on the triangular and honeycomb lattices decorated with a single hexagon per site. In certain parameter ranges, the Hubbard model maps to a quantum compass model on those lattices. On the triangular lattice, the compass model exhibits collinear stripe antiferromagnetism, implying $d$-density wave charge order in the original Hubbard model. On the honeycomb lattice, the compass model has a unique, quantum disordered ground state that transforms nontrivially under lattice reflection. The ground state of the Hubbard model on the decorated honeycomb lattice is thus a 2D fermionic symmetry-protected topological phase.
This state -- protected by time-reversal and reflection symmetries -- cannot be connected adiabatically to a free-fermion topological phase.
\end{abstract}

\pacs{71.10.Fd, 71.27.+a, 75.10.Jm}

\maketitle

\textit{Introduction.} The discovery of topological band insulators (TBI) of noninteracting electrons in certain strongly spin-orbit coupled semiconductors is one of the most important advances of the last decade in condensed matter physics~\cite{qi2010}. 
TBI are an example of symmetry-protected topological phases (SPT)~\cite{senthil2015}, which are devoid of intrinsic topological order such as that found in fractional quantum Hall systems, but possess protected edge/surface states with exotic characteristics. A major focus of current research is to discover interacting SPT phases that cannot be adiabatically deformed into noninteracting TBI.
While progress has been made in the classification~\cite{gu2009,pollmann2012,chen2011,chen2013,chen2012} and theoretical realization in model Hamiltonians~\cite{levin2012,senthil2013,regnault2013,geraedts2013,burnell2014,geraedts2014} of SPT phases of bosons, much less is known about SPT phases of fermions, which are relevant for electrons in solids. Although recent theories suggest that fermionic SPT phases distinct from free-fermion TBI should exist in principle~\cite{gu2014,wang2014}, apart from the special case of one spatial dimension (1D) there has been no explicit realization of a fermionic SPT as the ground state of a microscopic model Hamiltonian. In this paper, we provide evidence that a 2D fermionic SPT protected by time-reversal and reflection symmetries and distinct from a free-fermion TBI can be realized as the ground state of a simple Hubbard model for spin-1/2 electrons on a decorated honeycomb lattice.

\begin{figure}[t]
 \includegraphics[width=\columnwidth]{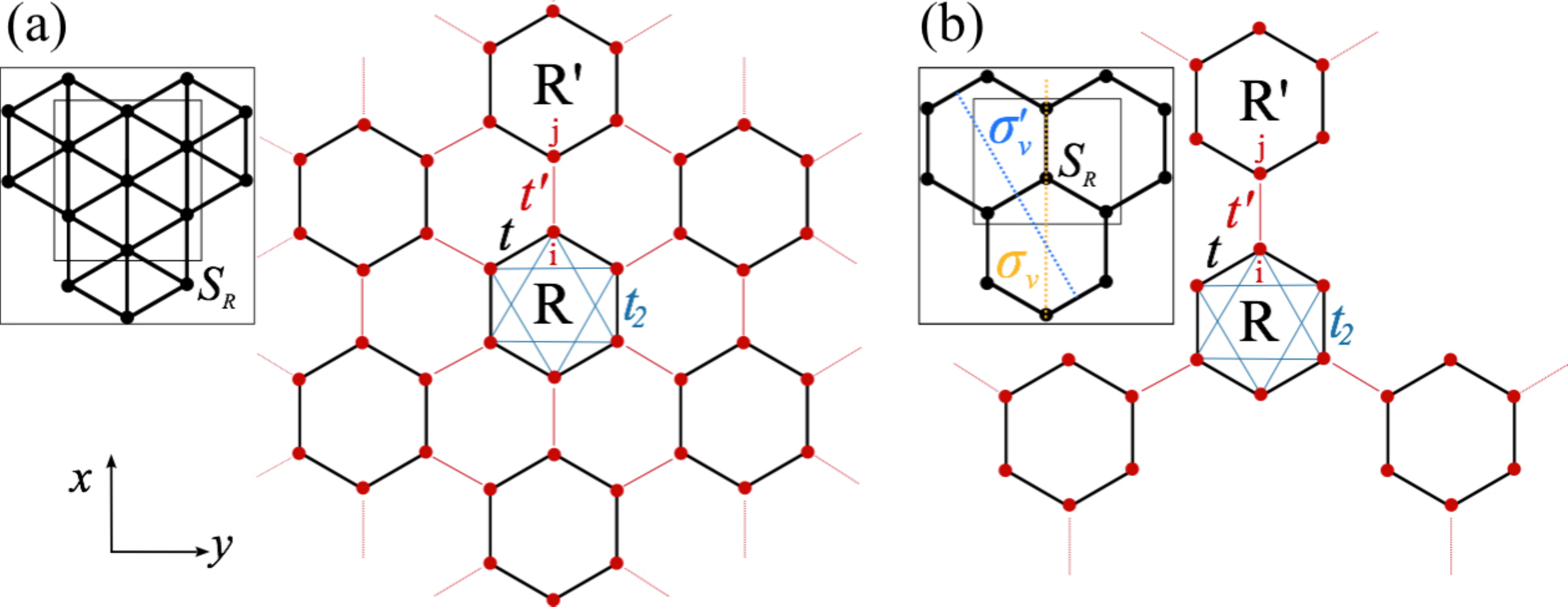}
\caption{Hubbard model on (a) the decorated triangular lattice; (b)  the decorated honeycomb lattice. Each site of the undecorated lattice is replaced by a six-site hexagon with intra-hexagon nearest-neighbor hopping $t$ and next-nearest-neighbor hopping $t_2$; the hexagons are connected by inter-hexagon nearest-neighbor hopping $t'$.}
\label{fig_1}
\end{figure}
\textit{Fermionic Hubbard model.} We consider the Hubbard model for spin-1/2 fermions hopping on the decorated triangular and honeycomb lattices (Fig.~\ref{fig_1}), where each site $\b{R}$ of the original triangular and honeycomb lattices is decorated by a single hexagon. Fermions hop within each hexagon with a nearest-neighbor amplitude $t$ and a next-nearest-neighbor amplitude $t_2$, and interact via an on-site repulsion $U>0$; hopping between the hexagons proceeds with an amplitude $t'$. The local Hamiltonian $\mathcal{H}_{\hexagon}^\b{R}$ for a hexagon on site $\b{R}$ is
\begin{align}\label{eq: local-Hamiltonian}
\mathcal{H}_{\hexagon}^\b{R}=&-\sum_\sigma\sum^6_{i=1}
\left(tc^{\dag}_{\b{R}i\sigma} c_{\b{R},i+1,\sigma}^{\phantom{\dag}}+t_2c^{\dag}_{\b{R}i\sigma} c_{\b{R},i+2,\sigma}^{\phantom{\dag}}   + \mathrm{H.c.} \right)\nonumber\\
&+U\sum_{i=1}^6 n_{\b{R}i\uparrow}n_{\b{R}i\downarrow},
\end{align}
where $c_{\b{R}i\sigma}^{\dag}$ ($c_{\b{R}i\sigma}^{\phantom{\dag}} $) creates (annihilates) a fermion of spin $\sigma$ on the $i$th vertex of the hexagon at $\b{R}$, and $n_{\b{R}i\sigma}\equiv c_{\b{R}i\sigma}^{\dag}c_{\b{R}i\sigma}^{\phantom{\dag}}$ is the fermion number operator. We define $c_{\b{R},i+6,\sigma}^{\phantom{\dag}} \equiv c_{\b{R}i\sigma}^{\phantom{\dag}} $, corresponding to periodic boundary conditions within the hexagon. We study the model at half filling with six fermions on each hexagon. The full Hamiltonian on the decorated lattice is then
\begin{align}
\label{eq: Fermionic-Hamiltonian}
\mathcal{H} &= \sum_{\textbf{R}} \mathcal{H}^{\textbf{R}}_{\hexagon} - t' \sum_{\langle \textbf{R}i,\textbf{R}'j\rangle,\sigma} \left( c^{\dag}_{\textbf{R}i\sigma} c_{\textbf{R}'j,\sigma}^{\phantom{\dag}}  + \mathrm{H.c.} \right),
\end{align}
where the sum in the second term runs over pairs of nearest-neighbor hexagons located at $\textbf{R},\textbf{R}'$, and $i,j$ are nearest-neighbor sites on the two adjacent hexagons (Fig.~\ref{fig_1}). The Hamiltonian possesses $SU(2)$ spin rotation symmetry, time-reversal ($T$) symmetry, and the $C_{6v}$ point group symmetry of the triangular Bravais lattice.

We wish to investigate the ground state properties of Eq. (\ref{eq: Fermionic-Hamiltonian}) in the limit of weakly coupled hexagons $t'\rightarrow 0$. For $t'=0$, the hexagons are decoupled and the many-body ground state is simply the product of the ground states of isolated hexagons. As shown previously~\cite{Moebius}, for $t_2 > t$ in a certain range of $U$, the ground state of an isolated hexagon is doubly degenerate and transforms as the $E_2$ irreducible representation of $C_{6v}$, whose two components have $d_{x^2-y^2}$ and $d_{xy}$ symmetry, respectively~\cite{sigrist1991}. This ground state doublet is separated from the excited states by a finite energy gap $\Delta$. The ground state of (\ref{eq: Fermionic-Hamiltonian}) for $t'=0$ thus has a macroscopic degeneracy of $2^N$, where $N$ is the total number of sites of the (undecorated) triangular or honeycomb lattice. Our strategy is to lift this macroscopic degeneracy by weakly coupling these strongly correlated hexagons with a nonzero infinitesimal $t'$, in the hope of uncovering interesting ground states for the full Hamiltonian (\ref{eq: Fermionic-Hamiltonian}). The rest of the paper assumes the values $U/t = 8$ and $t_2/t = 1.8$, corresponding to the $E_2$ ground state for each isolated hexagon~\cite{Moebius}.

The doubly degenerate $E_2$ ground states of an isolated hexagon at $\b{R}$ define a pseudospin-1/2 degree of freedom $\b{S}_\b{R}$, with $S_\b{R}^z=\pm 1/2$ corresponding to the complex linear combination $d_{xy}\pm i d_{x^2-y^2}$. This pseudospin thus can be interpreted as an orbital degree of freedom. In the limit $t'/\Delta\ll 1$, the macroscopic degeneracy of the decoupled hexagon problem is lifted by virtual hopping processes between the hexagons, and an effective Hamiltonian for the degenerate ground subspace can be derived by perturbation theory in powers of $t'/\Delta$. This effective Hamiltonian becomes a spin model for the pseudospin-1/2 degrees of freedom $\b{S}_\b{R}$ and, to leading order, is given by the quantum compass model~\cite{SuppMat},
\begin{align}
\label{eq: Spin-Hamiltonian}
\mathcal{H}_{\mathrm{S}} & =  \frac{J}{2} \sum_{\langle \textbf{R}\textbf{R}'\rangle} [\textbf{S}_\textbf{R} \cdot (\textbf{R}-\textbf{R}')] [\textbf{S}_{\textbf{R}'}\cdot (\textbf{R}-\textbf{R}')] ,
\end{align}
where the effective exchange coupling $J\sim (t')^2/\Delta > 0$ is antiferromagnetic and the sum is over pairs of nearest-neighbor sites on the (undecorated) triangular and honeycomb lattice. From the point of view of ground state properties, the mapping from the fermionic Hubbard model (\ref{eq: Fermionic-Hamiltonian}) to the effective spin Hamiltonian (\ref{eq: Spin-Hamiltonian}) is asymptotically exact in the $t'\rightarrow 0$ limit. 
Further, apart from the overall energy scale, Hamiltonian~(\ref{eq: Spin-Hamiltonian}) is independent of the exact values of the parameters $t_2/t$ and $U/t$ so long as the individual hexagons are in the $E_2$ phase and $t'/\Delta\ll 1$.
Our approach is to solve the spin model by exact diagonalization (ED) and determine the exact ground state for the fermionic problem in this limit~\cite{SuppMat}. The physical symmetries of the fermionic model are implemented in unusual ways in the spin model:
While the Hubbard model (\ref{eq: Fermionic-Hamiltonian}) exhibits spin $SU(2)$ symmetry for the fermions, the effective spin model (\ref{eq: Spin-Hamiltonian}) does not preserve either pseudospin $SU(2)$ or $U(1)$ symmetry.
$T$ symmetry only flips the sign of the out-of-plane component $S_\b{R}^z$ in Eq. (\ref{eq: Spin-Hamiltonian}), but not the in-plane components $S_\b{R}^x,S_\b{R}^y$. Furthermore, $T^2 = 1$ due to the even number of fermions on each hexagon. A $C_6$ spatial rotation in the fermionic model is implemented in the spin model as a simultaneous in-plane rotation of the lattice and the pseudospin operators. Apart from its connection to our fermionic problem, a solution of the compass model~(\ref{eq: Spin-Hamiltonian}) on the triangular and honeycomb lattices is interesting in its own right, given the relevance of this model to a host of physical systems ranging from spin-orbit coupled Mott insulators to ultracold atomic gases~\cite{CompassRevMod}.


\begin{figure}[t]
 \includegraphics[width=\columnwidth]{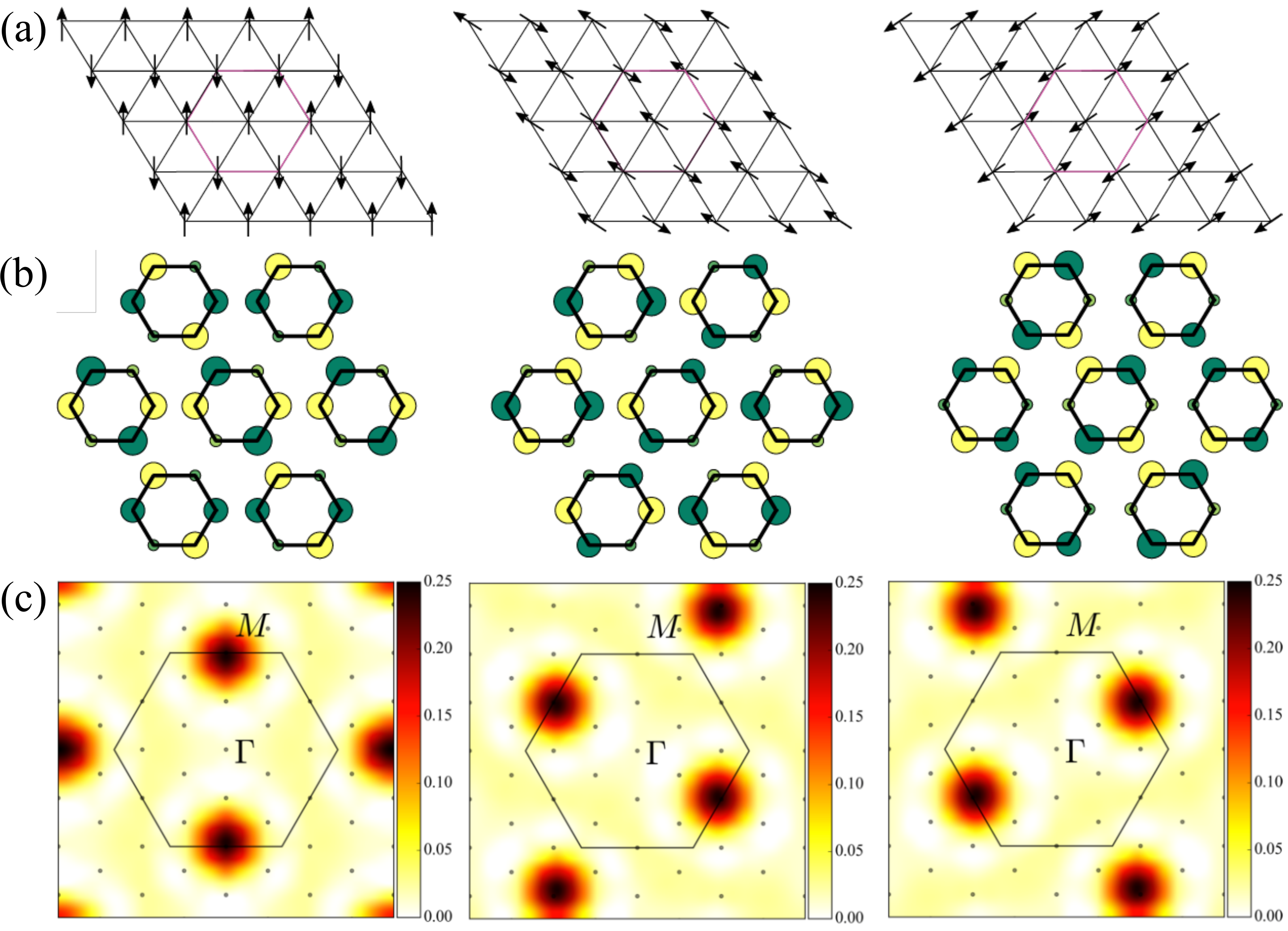}
\caption{(a) Classical ground states of the compass model \eqref{eq: Spin-Hamiltonian} on the triangular lattice. (b) Schematics of three different degenerate fermionic quantum ground states of the Hubbard model \eqref{eq: Fermionic-Hamiltonian}. The size of the circles at each site corresponds to the deviation of electron density from half filling. Yellow colors correspond to a surplus and green colors to a deficit.
(c) In-plane static structure factors $S^{\parallel}_{N=16}(\textbf{Q})$ of the corresponding ground states in the compass model.
The results were obtained by interpolating data on the available momentum points (indicated by black dots).
}
\label{fig_2}
\end{figure}

\textit{Triangular lattice.} Previous studies have considered the compass model~(\ref{eq: Spin-Hamiltonian}) on the triangular lattice in the semiclassical limit of large spins. The classical ground states were found to be six-fold degenerate and correspond to collinear stripe in-plane antiferromagnetic (AF) order as depicted in Fig.~\ref{fig_2}(a)~\cite{mostovoy2002,zhao2008,Wu120Compass}.  
Here we study whether the classical orders are stable against quantum fluctuations by solving \eqref{eq: Spin-Hamiltonian} with ED on $N$-site clusters ($N = 16,20,24,32$) under periodic boundary conditions. The cluster shapes are chosen to be compatible with the classical ordering patterns while preserving a maximal amount of point group symmetries~\cite{SuppMat}.
We begin by studying the static spin structure factor:
\begin{align}
S^{\alpha\beta}_N(\textbf{Q}) = \frac{1}{N^2} \sum_{\b{R}\b{R}'} e^{i\textbf{Q}\cdot(\textbf{R} - \textbf{R}')} \left\langle S^\alpha_\b{R} S^\beta_{\b{R}'}\right\rangle.
\end{align}
For all the cluster sizes under study, the in-plane component $S^{\parallel}_N(\textbf{Q})=S^{xx}_N(\b{Q})+S^{yy}_N(\b{Q})$ always peaks at the $M$ points [Fig.~\ref{fig_1}(c)] and dominates over the out-of-plane component $S^{zz}_N(\b{Q})$.
In addition, examining the individual components of $S^{\parallel}_N(\textbf{Q})$ reveals that the spins are mainly aligned parallel to the ordering vector $\textbf{Q}$.
These results indicate the ground state indeed exhibits the type of magnetic order expected classically.
In particular, finite-size scaling with $S^{\parallel}_N(M) = m^2_{\infty} + \frac{\alpha}{N} + \mathcal{O}(\frac{1}{N^2})$ renders a finite, positive value of $m_\infty \simeq 0.28$,
which is reduced by roughly $44\%$ from its classical value by quantum fluctuations.
We also note that the energy splitting between the two states lowest in energy decreases exponentially with cluster size as $e^{-\sqrt{N}/\xi}$, indicative of ground state degeneracy. The (six-fold) degenerate ground states are separated from the rest of the spectrum by a finite excitation gap extrapolated to be larger than $0.5J$ in the thermodynamic limit~\cite{SuppMat}.

Using the mapping between pseudospins and $E_2$ states of the fermionic hexagonal plaquettes, one can infer the nature of the ground states of the 2D Hubbard model (\ref{eq: Fermionic-Hamiltonian}) from the magnetically ordered ground states of the pseudospin Hamiltonian (\ref{eq: Spin-Hamiltonian}).
If out-of-plane ferromagnetic pseudospin order were to occur, this would correspond to a translationally invariant but $T$-breaking state of fermions with uniform $d_{xy}\pm id_{x^2-y^2}$ order. This state would also break rotational symmetry spontaneously due to the nontrivial $e^{\pm 2\pi i/3}$ eigenvalue of the $S^z_\b{R}$ eigenstates under $C_6$ rotations. However, the pseudospin in-plane AF order found here does not break the physical $T$ symmetry of the original fermion problem, as in-plane pseudospin components are even under $T$. Indeed, the state of a single hexagon at $\b{R}$ with in-plane pseudospin forming an angle $\phi_\b{R}$ with the $x$ axis is given by $\left|\phi_\b{R}\right\rangle=(\left|\uparrow\right\rangle+e^{i\phi_\b{R}}\left|\downarrow\right\rangle)/\sqrt{2}$ in the $S^z_\b{R}$ basis. This state corresponds to the real, and thus $T$-invariant, linear combination $\cos(\phi_\b{R}/2)d_{xy}+\sin(\phi_\b{R}/2)d_{x^2-y^2}$. To characterize how translation symmetry is broken, we calculate the density deviation from half filling $\delta n_{\b{R}i}(\phi)\equiv\left\langle\phi_\b{R}\right|\sum_\sigma n_{\b{R}i\sigma}-1\left|\phi_\b{R}\right\rangle$ for each of the six classical ground states, in which the pseudospin angles are $\phi_\b{R}=(2m+1)\pi/6$, $m=0,\ldots,5$.
As depicted schematically in Fig.~\ref{fig_2}(b), the resulting charge order in the fermion problem corresponds to a $T$-invariant $d$-density wave charge order~\cite{nayak2000} that doubles the unit cell. 

\begin{figure}[t]
\includegraphics[width=\columnwidth]{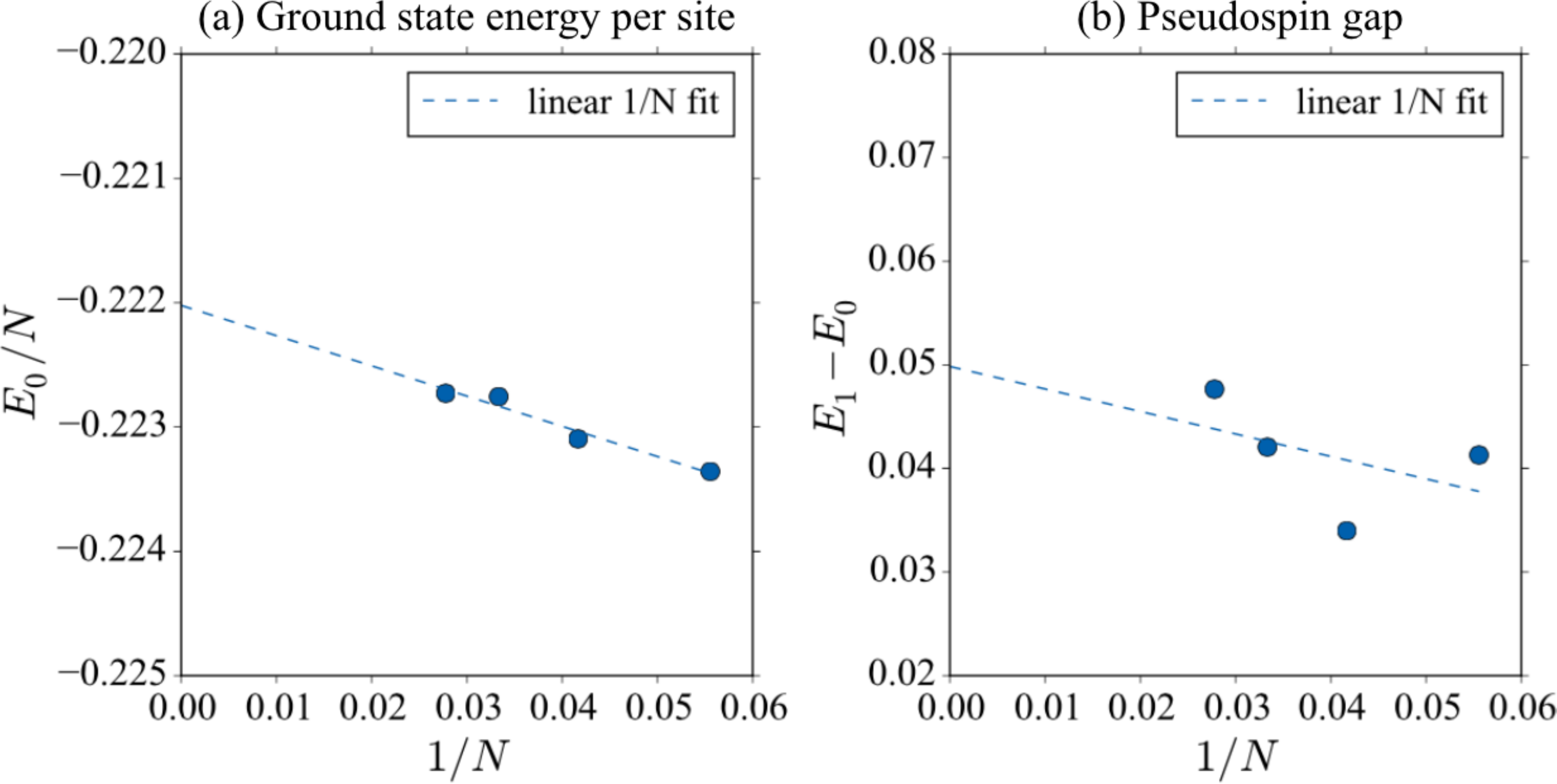}
\caption{
Exact diagonalization energy spectra of the compass model~\eqref{eq: Spin-Hamiltonian} on the honeycomb lattice:
(a) Ground-state energy per lattice site $E_0/N$. All ground states reside in the zero-momentum sector. Linear $1/N$ extrapolation yields $E_0/N\sim -0.222J$ in the thermodynamic limit.
(b) Energy gap above the ground state. Linear $1/N$ extrapolation yields a finite pseudospin gap $\sim 0.05J$ in the thermodynamic limit.
}
\label{fig_3}
\end{figure}

\textit{Honeycomb lattice.} Semiclassical studies of the compass model on the honeycomb lattice have established the existence of a macroscopic number of classical ground states~\cite{zhao2008,Wu120Compass,Nasu120Compass}; the model is thus frustrated. Whether and, if so, how the quantum model for spin-1/2 degrees of freedom orders at zero temperature is still under debate. Linear spin-wave theories predict that quantum order-by-disorder effects favor in-plane N\'eel order~\cite{zhao2008} or the so-called six-site plaquette order~\cite{Wu120Compass}.
An earlier ED study on clusters up to $N=24$ sites~\cite{Nasu120Compass} finds a quantum disordered ground state with gapless excitations,
although finite-size extrapolation cannot distinguish between gapless excited states and degenerate ground states. 
Furthermore, certain cluster geometries employed in that study frustrate the six-site plaquette order and thus introduce a bias.
A recent tensor network study~\cite{TNSorder} suggests that the ground state develops the six-site plaquette order. Tensor network algorithms have the advantage over ED that they can be formulated directly in the thermodynamic limit, but are intrinsically variational as one assumes that the ground-state wave function can be expressed as a network of local tensors defined on each lattice site. Here we use ED to study the $120^{\circ}$ quantum compass model on the honeycomb lattice, as in Ref.~\cite{Nasu120Compass}, but with larger cluster sizes ($N=18,24,30,36$) and geometries that support both the N\'eel and six-site plaquette orders~\cite{SuppMat}.
While not all the clusters we consider preserve the full $C_{6v}$ point group, they all contain a $C_{2v}$ subgroup with a $C_2$ rotation axis and two mirror planes $\sigma_v,\sigma_v'$ [see Fig. 1(b)]. 
Without pseudospin $SU(2)$ and $U(1)$ symmetries in Eq. (\ref{eq: Spin-Hamiltonian}), the $N=36$ calculation using translation symmetry corresponds to a Hamiltonian matrix of $\sim 3.8\times 10^9$ basis states.

Figure \ref{fig_3} shows the finite-size scaling of the ground state energy and the many-body gap to the first excited state. In spin-wave theory, the ground-state energy per lattice site $E_{0}/N$ is $-0.225 J$. The previous ED study with $N \leq 24$ reports $E_0/N = -0.215 J$ and a vanishing pseudospin gap~\cite{Nasu120Compass}. 
For clusters of the same size, we obtain a lower ground-state energy due to the higher spatial symmetry of our clusters, which can accommodate both the N\'eel and six-site plaquette orders. In addition, while the gap decreases rapidly with increasing $N$ for $N < 18$, it tends to saturate above $N = 18$. 
A linear $1/N$ extrapolation of our ED results with $N \geq 18$ yields $E_0/N = -0.222 J$ (lower than the tensor-network result of $-0.148J$~\cite{TNSorder}) and a finite excitation gap $\sim 0.05 J$ in the thermodynamic limit [Fig. \ref{fig_3}]. 

\begin{figure}[t]
 \includegraphics[width=\columnwidth]{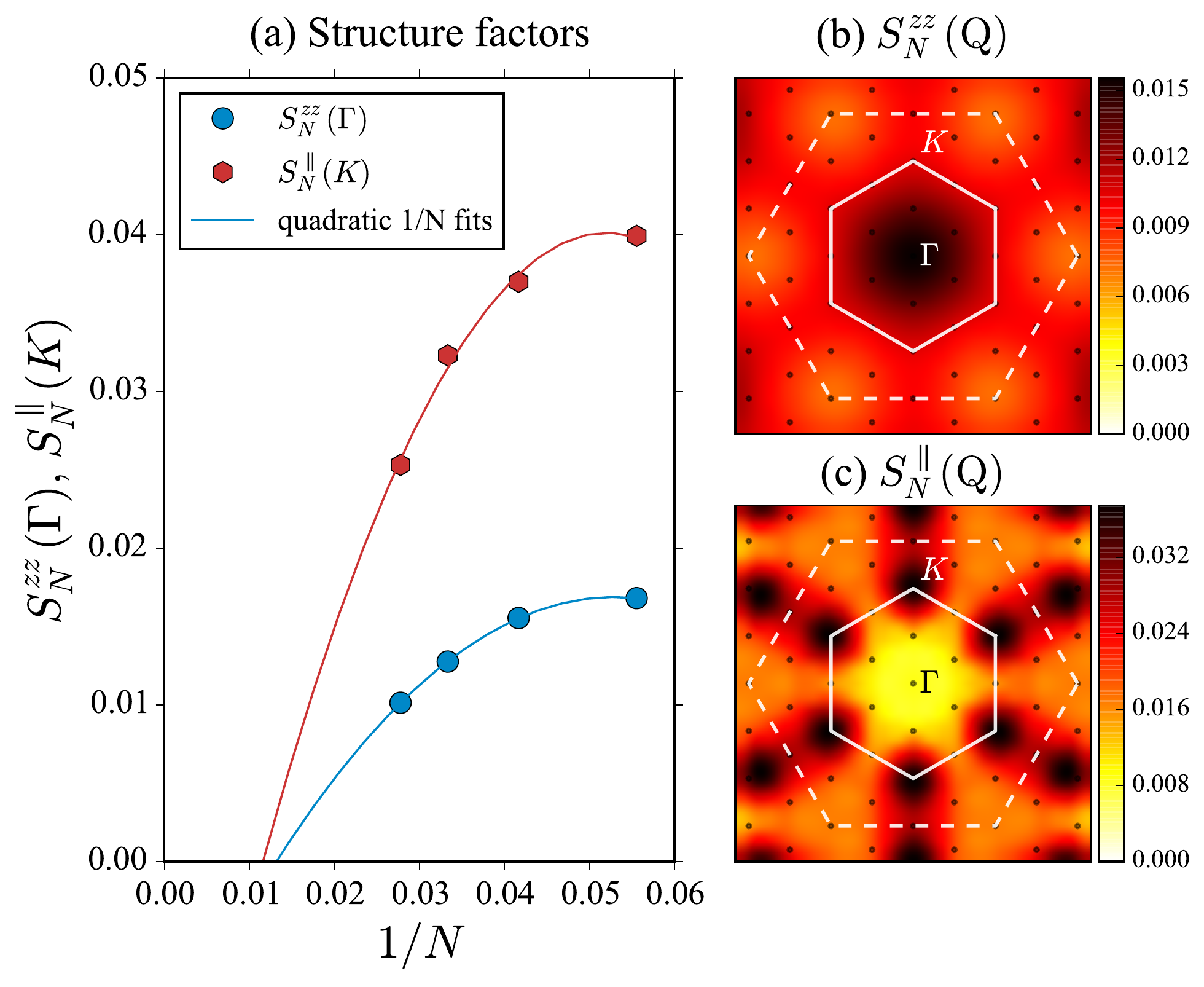}
\caption{
Static structure factors obtained by exact diagonalization of the compass model \eqref{eq: Spin-Hamiltonian} on the honeycomb lattice:
(a) Scalings of $S^{zz}(\textbf{Q}=\Gamma)$ and $S^{\parallel}(\textbf{Q}=K)$.
The results show the absence of N\'eel and six-site plaquette orders, respectively.
(b)-(c) Maps of $S^{zz}_{N=24}(\textbf{Q})$ and $S^{\parallel}_{N=24}(\textbf{Q})$ obtained by interpolating data on the available momentum points (indicated by black dots).
The first and second Brillouin zones are denoted by the solid and dashed white hexagons, respectively.
}
\label{fig_4}
\end{figure}

For all clusters, the out-of-plane structure factor $S^{zz}_N(\textbf{Q})$ peaks at the $\Gamma$ point [Fig.~\ref{fig_4}(b)], corresponding to short-range N\'eel order (or ferromagnetic order after a sublattice basis rotation).
However, $S^{zz}_N(\Gamma)$ decreases faster than $1/N$.
Both quadratic $1/N$ scaling for $N \geq 18 $ and linear $1/N$ scaling for $N \geq 24$ lead to an extrapolated $S^{zz}_{N=\infty}(\Gamma) < 0$, which thereby rules out long-range N\'eel order in the thermodynamic limit [Fig.~\ref{fig_4}(a)].
On the other hand, the in-plane structure factor $S^{\parallel}_N(\textbf{Q})$ peaks at the $K$ points [Fig.~\ref{fig_4}(c)], reminiscent of a short-range six-site plaquette order.
$S^{\parallel}_{N=\infty}(K)$, however, also extrapolates to a negative value in the thermodynamic limit [Fig.~\ref{fig_4}(a)].
The pseudospin correlation length is estimated to be less than one unit-cell length based on the structure factor~\cite{sandvik2010}.
Our results thus suggest that long-range six-site plaquette order is also absent, and the ground state of the $120^{\circ}$ compass model on the honeycomb lattice is a quantum pseudospin liquid.
Because the honeycomb lattice has two sites per unit cell, according to the generalized Hastings-Oshikawa-Lieb-Schultz-Mattis theorem~\cite{Watanabe15}, a gapped ground state without symmetry breaking does not imply intrinsic topological order with long-range entanglement~\cite{Kitaev_PRL_2006, Levin_PRL_2006}. 
Since we find a unique ground state on the torus (as periodic boundary conditions are employed in both directions), we conclude that the ground state of the compass model --- and thus that of the fermionic Hubbard model --- are short-range entangled.

A gapped, symmetric, short-range entangled ground state for the fermionic Hubbard model (\ref{eq: Fermionic-Hamiltonian}) is either adiabatically connected to a free-fermion trivial band insulator, a free-fermion TBI, or a fermionic SPT phase distinct from the latter. To distinguish between the three, we classify the ground states of the compass model (\ref{eq: Spin-Hamiltonian}) according to irreducible representations of the cluster point group $C_{2v}$. Indeed, the ground state of a $T$-invariant free-fermion band insulator, trivial or topological, must transform according to the identity representation of the point group~\cite{FMI}. For an adiabatic path that preserves $T$ and point group symmetries, a ground state belonging to a nontrivial 1D point-group representation cannot be adiabatically connected to a free-fermion ground state and is thus a fermionic SPT protected by those symmetries. On clusters with $N/2$ even, the ground state of the compass model --- and thus that of the corresponding Hubbard model --- belongs to the identity representation of $C_{2v}$ and cannot be distinguished from a band insulator by its transformation properties under point group symmetries alone. With $N/2$ odd, the ground state is odd under the reflections $\sigma_v$ and $\sigma_v'$ indicated by the dotted lines in Fig.~\ref{fig_1}(b) and even under $C_2$ rotation: It thus transforms according to the $A_2$ representation of the $C_{2v}$ point group and cannot be adiabatically connected to a band insulator. (See Ref.~\cite{Kimchi_PNAS2013, Ware_PRB2015, TNSA2} for a tensor-network construction of a state with similar transformation properties.) In the thermodynamic limit $N\to \infty$, a band insulator would transform trivially under all point group operations for both $N/2$ odd and $N/2$ even, {\it i.e.}, independently of how the thermodynamic limit is approached. Our results therefore suggest that the ground state of the fermionic Hubbard model~(\ref{eq: Fermionic-Hamiltonian}) on the decorated honeycomb lattice realizes a 2D fermionic SPT distinct from a free-fermion TBI and protected by $T$ and $C_{2v}$ symmetries. Our results hold in the asymptotic limit $t'\rightarrow 0$ for a finite range of $U$ and $t_2$ so long as individual hexagons are in the $E_2$ phase~\cite{Moebius}.

\textit{Concluding remarks.} Our numerical ED results suggest that a simple 2D Hubbard model of spin-1/2 electrons exhibits $d$-density wave charge order on the decorated triangular lattice but produces an interacting fermionic SPT distinct from a free-fermion TBI on the decorated honeycomb lattice. Given the relative simplicity of our model Hamiltonian, we expect our findings to assist in the experimental search for SPT phases of electrons in real materials. Our study is conceptually similar to that of the checkerboard Hubbard model on the square lattice, in which isolated square plaquettes with strongly correlated ground states transforming as the $B_1$ ($d_{x^2-y^2}$) representation of $C_{4v}$ couple weakly and form an exotic ground state for the entire system, the $d$-Mott insulator~\cite{yao2007,FMI,PlaquetteDMFT}, which also belongs to the $B_1$ representation. 
While this is also strictly speaking a fermionic SPT (protected by $T$ and $C_{4v}$ symmetries), it is adiabatically connected to a product state of decoupled plaquettes occupying the sites of a decorated square lattice. Because $C_{4v}$ acts on such plaquettes as an on-site symmetry of this lattice, the $d$-Mott insulator can be viewed as a ``stack'' of 0D SPTs protected by on-site symmetries ($T$ and $C_{4v}$) and the translation symmetry of the decorated lattice, i.e., as a weak SPT in the sense of Refs. \onlinecite{chen2013, cheng2015}.
By contrast, the ground state found here cannot be reduced to a product state of $N$ hexagonal plaquettes in the $A_2$ representation, as such a product state would transform trivially under $\sigma_v$ with eigenvalue $(-1)^N=1$ since $N$ is even on the honeycomb lattice, irrespective of whether $N/2$ is even or odd. Since the protecting $C_{2v}$ symmetry of our state generally does not act as an on-site symmetry, except in the (excluded) case of a product state of 0D SPTs, we believe our state cannot be a weak SPT.

\textit{Acknowledgements.} 
The authors acknowledge discussions with B. Bauer, K. Penc, H.-C. Jiang, and T. F. Seman.
C.C.C. was supported by the Aneesur Rahman Postdoctoral Fellowship at Argonne National Laboratory, operated by the U.S. Department of Energy (DOE) Contract No. DE-AC02-06CH11357.
L.M. and R.C. were supported by the U. S. DOE Contract No. DE-FG02-05ER46201.
J.M. was supported by NSERC grant \#RGPIN-2014-4608, the Canada Research Chair Program (CRC), the Canadian Institute for Advanced Research (CIFAR), and the University of Alberta.
This research used resources of the National Energy Research Scientific Computing Center, supported by the U.S. DOE Contract No. DE-AC02-05CH11231.

\bibliography{lit}

\end{document}


\preprint{}

\title{Supplemental Material: Fermionic Symmetry-Protected Topological Phase in a Two-Dimensional Hubbard Model}

\author{Cheng-Chien Chen}
\affiliation{Advanced Photon Source, Argonne National Laboratory, Argonne, Illinois 60439, USA}
\affiliation{Department of Physics, University of Alabama at Birmingham, Birmingham, Alabama 35294, USA}
\author{Lukas Muechler}
\affiliation{Department of Chemistry, Princeton University, Princeton, New Jersey 08544, USA}
\author{Roberto Car}
\affiliation{Department of Chemistry, Princeton University, Princeton, New Jersey 08544, USA}
\author{Titus Neupert}
\affiliation{Princeton Center for Theoretical Science, Princeton University, Princeton, New Jersey 08544, USA}
\author{Joseph Maciejko}
\affiliation{Department of Physics, University of Alberta, Edmonton, Alberta T6G 2E1, Canada}
\affiliation{Theoretical Physics Institute, University of Alberta, Edmonton, Alberta T6G 2E1, Canada}
\affiliation{Canadian Institute for Advanced Research, Toronto, Ontario M5G 1Z8, Canada}

\date\today

\maketitle

\appendix
\onecolumngrid
\vspace{\columnsep}

In this Supplemental Material, we give further technical details concerning the derivation of the effective pseudospin Hamiltonian [Eq. (3) in the main text] as well as the numerical exact diagonalization procedure.

\section{I. Derivation of the Compass Model from the Hubbard Model}
The $4 \times 4$ Hamiltonian for the hopping between site $i$ and $j$ of two adjacent hexagons $A$ and $B$ in second order perturbation theory (PT2) is of the form 

\begin{align}\label{Perturbation}
H_{\lambda_{A_j}\lambda_{B_i},\tilde{\lambda}_{A_j} \tilde{\lambda}_{B_i} }
= (t')^2 \sum_{n_5,m_7 = 0}^{791}  \sum_{\sigma}
\frac{
 \braket{\lambda_{A_j}\lambda_{B_i}|c^{\dag}_{A_j\sigma} c_{B_i\sigma}^{\phantom{\dag}}|n_5,m_7}
\braket{n_5,m_7|c^{\dag}_{B_i\sigma} c_{A_j\sigma}^{\phantom{\dag}}|\tilde{\lambda}_{A_j} \tilde{\lambda}_{B_i} }
}
{2E_0 - E_{n_5} - E_{m_7}},
\end{align}
where $(B_i,A_j)$ correspond to neighboring lattice sites of adjacent hexagons and $\lambda = +,-$ labels the two $E_2$ states by their $\sigma_z$ eigenvalues. $E_{n_5}(E_{m_7})$ and $\ket{n_5} (\ket{m_7})$ are the energies and eigenstates of the five-(seven) electron problem on a single hexagon. $E_0$ is the energy of the doubly degenerate state $\ket{\pm}$ with six electrons on a hexagon. All matrix elements of this Hamiltonian can be calculated using the 792 energies and eigenvectors from exact diagonalization (ED). \\

The corresponding real space PT2-Hamiltonian on the lattice can then be expressed as
\begin{align}
\mathcal{H}^{(2)} & =   \sum_{<ij>} \sum_{\lambda_{A_j}\lambda_{B_i},\tilde{\lambda}_{A_j} \tilde{\lambda}_{B_i} } \ket{\lambda_{A_j}\lambda_{B_i}}   H^{ij}_{\lambda_{A_j}\lambda_{B_i},\tilde{\lambda}_{A_j} \tilde{\lambda}_{B_i} } \bra{\tilde{\lambda}_{A_j} \tilde{\lambda}_{B_i}},
\end{align}
where the sum runs over all nearest-neighbor links between adjacent hexagons.
This Hamiltonian can be expanded in a spin-basis $\sigma_{\mu} \otimes \sigma_{\nu}$ with the Pauli-matrices $\sigma_i$ and $ \sigma_0 \equiv \mathds{1}_{2\times2}$. Here, the $\ket{\pm}$ states on each hexagon are mapped to the $\sigma_z$ pseudospin-eigenstates.
Using the ED data, the pseudospin Hamiltonian takes on the form of a compass model with $J > 0$ for both the triangular and honeycomb lattices
\begin{align}
\mathcal{H}_{\mathrm{S}} & =  \frac{J}{2} \sum_{\langle \textbf{R},\textbf{R}'\rangle} [\textbf{S}_\textbf{R} \cdot (\textbf{R}-\textbf{R}')] \ [\textbf{S}_{\textbf{R}'}\cdot (\textbf{R}-\textbf{R}')].
\end{align}
$\textbf{S}_{\textbf{R}}= (\sigma^{\textbf{R}}_x,\sigma^{\textbf{R}}_y, \sigma^{\textbf{R}}_z)$ is the pseudospin-vector on site $\textbf{R}$ and the sum runs over pairs of nearest-neighbor pseudospins.

\section{II. Exact Diagonalization Calculations}
\begin{figure*}[t!]
\includegraphics[width=0.7\columnwidth]{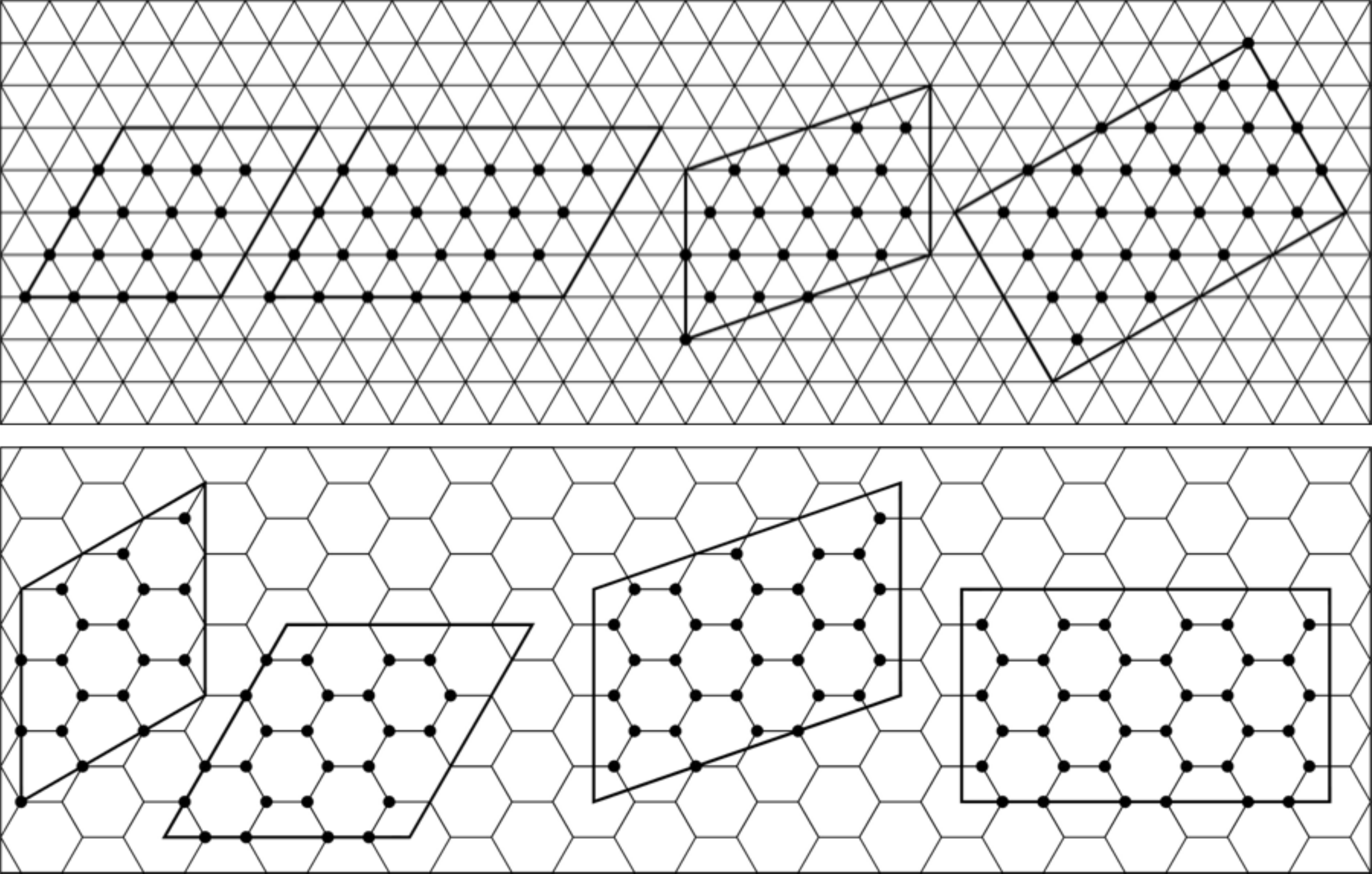}
\caption{
Clusters employed in the exact diagonalization calculations of the compass model on the triangular and honeycomb lattices. The black dots denote the choice of pseudospin sites.
}\label{fig_sm_lattice}
\end{figure*}

The ED calculations are performed with fully periodic boundary conditions using translational symmetry, where the Hamiltonian matrix is constructed on a basis of translation operator eigenstates. The employed clusters of size $N$ are shown in Fig.~\ref{fig_sm_lattice}; their geometries are chosen to be compatible with the classical magnetic ordering patterns while at the same time preserve a maximal amount of point-group symmetry operations. On the honeycomb lattice, all the clusters under study preserve a $C_{2v}$ subgroup. 
\\

On each cluster, we diagonalize the Hamiltonian matrix in all distinct momentum sectors to obtain the ground state. On the honeycomb lattice, the ground states on different clusters all reside in the zero-momentum sector. In this case, we further characterize the ground state according to its transformation properties under $C_{2v}$ point-group symmetry operations. In particular, with $N/2$ even, the ground state belongs to the identity representation. With $N/2$ odd, the ground state is odd under $\sigma_v$ and $\sigma_v'$ lattice reflections and even under $C_2$ rotation. 
\\

Since the compass model does not preserve the pseudospin $SU(2)$ and $U(1)$ symmetries, the computational complexity is substantially increased compared for example to the Heisenberg model on the same cluster. The $N = 36$ calculation for the compass model on the honeycomb lattice using only translation symmetry corresponds to a Hamiltonian matrix of $\sim 3.8 \times 10^9$ basis states. We use distributed memory parallelization based on the message passing interface, as well as the PETSc~\cite{PETSC} and SLEPc~\cite{SLEPC} libraries to solve the large-scale sparse-matrix eigenvalue problem. The Krylov-Schur~\cite{Krylov_Schur_2001} algorithm chosen for the iterative matrix diagonalization is an improved variation of the Arnoldi method with optimized implicit restarting and ability to obtain multiple degenerate eigenstates.
\\

\begin{figure*}[h!]
\includegraphics[width=\columnwidth]{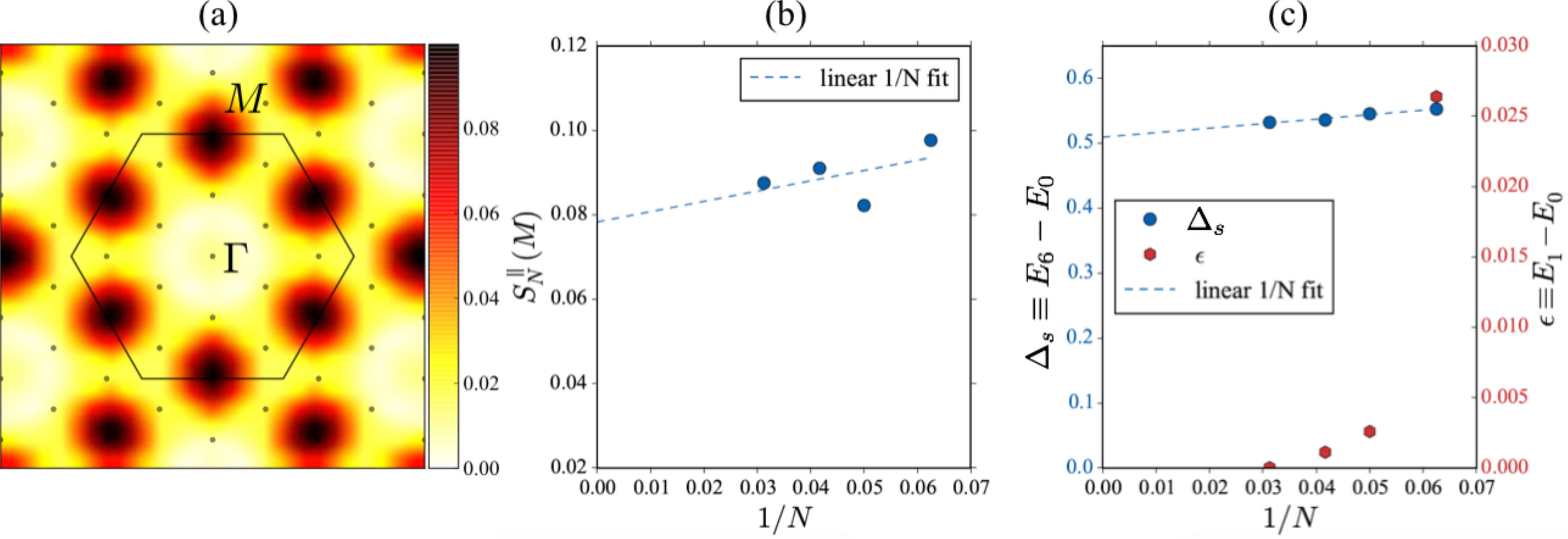}
\caption{
Exact diagonalization results for the compass model on the triangular lattice:
(a) The in-plane static structure factor $S^{\parallel}_{N=16}(\textbf{Q})$ for the ground state in the zero-momentum sector.
The false-color intensity map is obtained by extrapolating data on the available momentum points (black dots) in the Brillouin zone.
(b) Linear $1/N$ extrapolation of $S^{\parallel}_N(\textbf{Q}=M)$. (c) Energy splitting $\epsilon$ between the two states lowest in energy and the pseudospin excitation gap $\Delta_s$ above the six-fold (quasi-)degenerate ground states.  }\label{fig_sm_triangular}
\end{figure*}

Figure \ref{fig_sm_triangular} shows additional ED results for the compass model on the triangular lattice. The in-plane static structure factor $S^{\parallel}_{N=16}(\textbf{Q})$ for the zero-momentum ground state peaks at the $M$ points, related to collinear stripe antiferromagnetism [Fig.~\ref{fig_sm_triangular}(a)].  The presence of long-range pseudospin magnetism can be established by extrapolating $S^{\parallel}_{N}(\textbf{Q}=M)$ with $S^{\parallel}_N(M)=m^2_{\infty}+\frac{\alpha}{N} + {\cal{O}}(\frac{1}{N^2})$ [Fig.~\ref{fig_sm_triangular}(b)]. The extrapolation yields a positive finite $m^2_{\infty}$; the triangular-lattice quantum compass model thus develops long-range collinear stripe antiferromagnetic order. Finally, Fig.  \ref{fig_sm_triangular}(c) shows the energy splitting $\epsilon$ between the two states lowest in energy and the pseudospin excitation gap $\Delta_s$ above the six-fold (quasi-)degenerate ground states. While $\Delta_s$ decreases with $1/N$ linearly and extrapolates to a finite value $> 0.5 J$ in the thermodynamic limit, $\epsilon$ decreases as $e^{-\sqrt{N}/\xi}$, which is characteristic of ground-state degeneracy.

%
%
%
%

\bibliography{lit_SM}